\documentstyle[preprint,aps,epsfig]{revtex}
\begin{document}
\draft

\title{\large {\bf The relativistic continuum Hartree-Bogoliubov
description of charge-changing cross section for C, N, O and F
isotopes \\ }}

\author{J. Meng$^{1-3}$\thanks{e-mail: mengj@pku.edu.cn},
S.-G. Zhou$^{1-3}$ and I. Tanihata$^{4}$}
\address{${}^{1}$Department of Technical Physics, Peking University,
                 Beijing 100871, China}
\address{${}^{2}$Institute of Theoretical Physics, Chinese Academy of Sciences,
                 Beijing 100080, China}
\address{${}^{3}$Center of Theoretical Nuclear Physics, National Laboratory of
                 Heavy Ion Accelerator, Lanzhou 730000, China}
\address{${}^{4}$The Institute of Physical and Chemical Research (RIKEN),
                 Hirosawa 2-1, Wako-shi, Saitama 351-0198,  Japan}
\date{\today}
\maketitle

\begin{abstract}
The ground state properties including radii, density distribution
and one neutron separation energy for C, N, O and F isotopes up
to the neutron drip line are systematically studied by the fully
self-consistent microscopic Relativistic Continuum
Hartree-Bogoliubov (RCHB) theory. With the proton density
distribution thus obtained, the charge-changing cross sections
for C, N, O and F isotopes are calculated using the Glauber model.
Good agreement with the data has been achieved. The charge
changing cross sections change only slightly with the neutron
number except for proton-rich nuclei. Similar trends of variations
of proton radii and of charge changing cross sections for each
isotope chain is observed which implies that the proton density
plays important role in determining the charge-changing cross
sections.
\end{abstract}


\par
\narrowtext


Recent progresses in the accelerator and detection techniques all
around the world have made it possible to produce and study the
nuclei far away from the stability line --- so called ``EXOTIC
NUCLEI''. Based on the measurement of interaction cross section
with radioactive beams at relativistic energy, novel and entirely
unexpected features has appeared: e.g., the neutron halo and skin
as the rapid increase in the measured interaction cross-sections
in the neutron-rich light nuclei \cite{THH.85b,HJJ.95}.

Systematic investigation of interaction cross sections for an
isotope chain or an isotone chain can provide a good opportunity
to study the density distributions over a wide range of isospin
\cite{Suz.95,MTY.97}. However the contribution from proton and
neutron are coupled in the measurement of interaction cross
section. To draw conclusion on the differences in proton and
neutron density distributions definitely, a combined analysis of
the interaction cross section and other experiment on either
proton or neutron alone are necessary.

The charge-changing cross section which is the cross section for
all processes which result in a change of the atomic number for
the projectile can provide good opportunity for this purpose. In
Ref.\cite{Chu.00}, the total charge-changing cross section
$\sigma_{\rm cc}$ for the light stable and neutron-rich nuclei at
relativistic energy on a carbon target were measured. We will
study $\sigma_{\rm cc}$ theoretically by using the fully
self-consistent and microscopic relativistic continuum
Hartree-Bogoliubov (RCHB) theory and the Glauber Model in the
present letter.

The RCHB theory \cite{ME.98,MR.96,MR.98}, which is an extension of
the relativistic mean field (RMF) \cite{SW86,RE89,RI96} and the
Bogoliubov transformation in the coordinate representation, can
describe satisfactorily the ground state properties for nuclei
both near and far from the $\beta$-stability line and from light
to heavy or super heavy elements, as well as for the understanding
of pseudo-spin symmetry in finite nuclei
\cite{MSY.98,Meng993,Meng992,Meng00}. A remarkable success of the
RCHB theory is the self-consistent reproduction of the halo in
$^{11}$Li \cite{MR.96} and the prediction of giant halo
\cite{MR.98}. In combination with the Glauber model, the RCHB
theory successfully reproduces the interaction cross section in
Na isotopes \cite{MTY.97}. These successes encourage us to apply
the RCHB theory to calculate the charge changing cross section of
the C, N, O, F isotopes (ranging from the $\beta$-stability line
to the neutron drip line) on the target of $^{12}$C reported in
Ref.\cite{Chu.00}. With the density distribution provided by RCHB
theory, the total charge-changing cross section can be calculated
based on the Glauber model and compared with the data directly
\cite{Chu.00}, as was done in Ref.\cite{MTY.97} for the
interaction cross section. Since the theory used here is fully
microscopic and basically parameter free, we hope it provide us
more reliable information on both the proton and neutron
distribution.

The ground state properties of C, N, O and F isotopes up to
neutron drip line are studied first, including single neutron
separation energies, density distributions and radii. Then the
total charge-changing cross sections will be calculated from the
Glauber model with the density obtained from RCHB calculations.

The basic ansatz of the RMF theory is a Lagrangian density
whereby nucleons are described as Dirac particles which interact
via the exchange of various mesons (the scalar sigma ($\sigma$),
vector omega ($\omega$) and iso-vector vector rho ($\rho$)) and
the photon. The $\sigma$ and $\omega$ meson provide the
attractive and repulsive part of the nucleon-nucleon force,
respectively. The necessary isospin asymmetry is provided by the
$\rho$ meson. The scalar sigma meson moves in a self-interacting
field having cubic and quadratic terms with strengths $g_2$ and
$g_3$ respectively. The Lagrangian then consists of the free
baryon and meson parts and the interaction part with minimal
coupling, together with the nucleon mass $M$ and $m_\sigma$
($g_\sigma$), $m_\omega$ ($g_\omega$), and $m_\rho$ ($g_\rho$)
the masses (coupling constants) of the respective mesons:
\begin{equation}
 \begin{array}{rl}
 {\cal L} &= \bar \psi (i\rlap{/}\partial -M) \psi +
  \,{1\over2}\partial_\mu\sigma\partial^\mu\sigma-U(\sigma)
  -{1\over4}\Omega_{\mu\nu}\Omega^{\mu\nu}\\
  \ &+ {1\over2}m_\omega^2\omega_\mu\omega^\mu
  -{1\over4}{\vec R}_{\mu\nu}{\vec R}^{\mu\nu} +
  {1\over2}m_{\rho}^{2} \vec\rho_\mu\vec\rho^\mu
  -{1\over4}F_{\mu\nu}F^{\mu\nu} \\
  &-  g_{\sigma}\bar\psi \sigma \psi~
     -~g_{\omega}\bar\psi \rlap{/}\omega \psi~
     -~g_{\rho}  \bar\psi
      \rlap{/}\vec\rho
      \vec\tau \psi
     -~e \bar\psi \rlap{/}A \psi
 \end{array}
\end{equation}.

For the proper treatment of the pairing correlations and for
correct description of the scattering of Cooper pairs into the
continuum in a self-consistent way, one needs to extend the
present relativistic mean-field theory to the
RCHB\cite{ME.98,MR.96,MR.98}:
\begin{equation}
 \left(\begin{array}{cc}
  h-\lambda & \Delta \\
  -\Delta^* & -h^*+\lambda
 \end{array}\right)
 \left(\begin{array}{r}
  U \\ V\end{array}\right)_k~=~
 E_k\,\left(\begin{array}{r} U \\ V\end{array}\right)_k,
\label{RHB}
\end{equation}
$E_k$ is the quasi-particle energy,  the coefficients $U_k(r)$ and $V_k(r)$ are
four-dimensional Dirac spinors, and $h$ is the usual Dirac Hamiltonian
\begin{equation}
 h =  \left[ {\mbox{\boldmath$\alpha$}} \cdot {\bf p} +
      V( {\bf r} ) + \beta ( M + S ( {\bf r} ) ) \right],
 \label{h-field}
\end{equation}
with the vector and scalar potentials calculated from:
\begin{eqnarray}
\left\{
\begin{array}{lll}
   V( {\bf r} ) &=&
      g_\omega\rlap{/}\omega({\bf r})
         + g_\rho\rlap{/}\mbox{\boldmath$\rho$}({\bf r})\mbox{\boldmath$\tau$}
         + \displaystyle\frac{1}{2}e(1-\tau_3)\rlap{\,/}{\bf A}\mbox{\boldmath$\tau$}({\bf r}) , \\
   S( {\bf r} ) &=&
      g_\sigma \sigma({\bf r}). \\
\end{array}
\right.
\label{vaspot}
\end{eqnarray}
The chemical potential $\lambda$ is adjusted to the proper
particle number. The meson fields are determined as usual in a
self-consistent way from the Klein Gordon equations in {\it
no-sea}-approximation.

The pairing potential $\Delta$ in Eq. (\ref{RHB}) is given
by
\begin{equation}
\Delta_{ab}~=~\frac{1}{2}\sum_{cd} V^{pp}_{abcd} \kappa_{cd}
\label{gap}
\end{equation}
It is obtained from the pairing tensor $\kappa=U^*V^T$ and the
one-meson exchange interaction $V^{pp}_{abcd}$ in the
$pp$-channel. As in Ref. \cite{ME.98,MR.96,MR.98} $V^{pp}_{abcd}$
in Eq. (\ref{gap}) is the density dependent two-body force of
zero range:
\begin{equation}
 V(\mbox{\boldmath $r$}_1,\mbox{\boldmath $r$}_2)
 ~=~\frac{V_0 }{2}(1+P^\sigma)
 \delta(\mbox{\boldmath $r$}_1-\mbox{\boldmath$r$}_2)
 \left(1 - \rho(r)/\rho_0\right).
 \label{vpp}
\end{equation}

The ground state $|\Psi\rangle$ of the even particle system is
defined as the vacuum with respect to the quasi-particle:
$\beta_{\nu} |\Psi\rangle=0$, $|\Psi\rangle = \prod_\nu
\beta_{\nu} |-\rangle$, where $|-\rangle$ is the bare vacuum. For
odd system, the ground state can be correspondingly written as:
$|\Psi \rangle_{\mu} = \beta_{\mu}^\dagger\prod_{\nu \ne \mu}
\beta_{\nu} | - \rangle$, where $\mu$ is the level which is
blocked. The exchange of the quasiparticle creation operator
$\beta_{\mu}^\dagger$ with the corresponding annihilation
operator $\beta_{\mu}$ means the replacement of the column $\mu$
in the  $U$ and $V$ matrices by the corresponding column in the
matrices $V^*$, $U^*$ \cite{RS.80}.

The RCHB equations (\ref{RHB}) for zero range pairing forces are
a set of four coupled differential equations for the
quasi-particle Dirac spinors $U(r)$ and $V(r)$. They are solved
by the shooting method in a self-consistent way as \cite{ME.98}.
The detailed formalism and numerical techniques of the RCHB
theory can be found in Ref.\cite{ME.98} and the references
therein. In the present calculations, we follow the procedures in
Ref.\cite{ME.98,MR.98,MTY.97} and solve the RCHB equations in a
box with the size $R=20$ fm and a step size of 0.1 fm. The
parameter set NL-SH \cite{SNR.93} is used, which aims at
describing both the stable and exotic nuclei. The density
dependent $\delta$-force in the pairing channel with
$\rho_0=0.152$ fm$^{-3}$ is used and its strength $V_0$ is fixed
by the Gogny force as in Ref.\cite{ME.98}. The contribution from
continua is restricted within a cut-off energy $E_{cut}\sim
120$MeV.

\begin{figure}
\centerline{\epsfig{figure=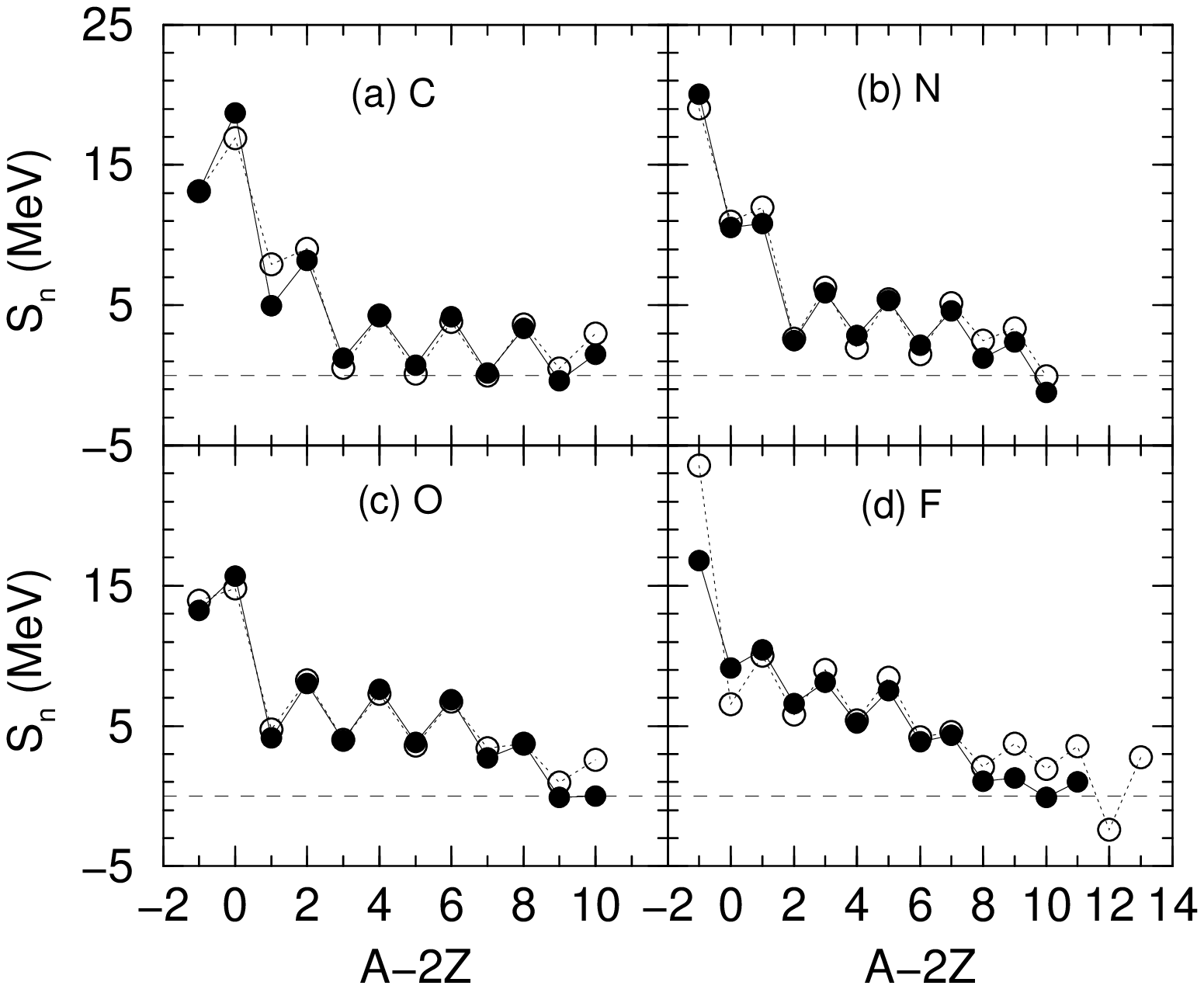,width=15.cm}}
 \vspace*{20pt}
 \caption{The one neutron separation energies $S_{\rm n}$
 for the nuclei $^{11-22}$C,$^{13-24}$N,$^{15-26}$O and
 $^{17-25}$F by RCHB theory(open circles) and their experimental
 counterparts (solid circles).}
 \label{fig.s}
\end{figure}

Systematic calculations with RCHB theory has been carried out for the C, N, O and F
isotopes. The one neutron separation energies $S_{\rm n}$ predicted by RCHB and their
experimental counterparts \cite{AUD.93} for the nuclei $^{11-22}$C, $^{13-24}$N,
$^{15-26}$O and $^{17-25}$F are given in Fig.\ref{fig.s} as open and solid circles
respectively. For carbon isotopes, the theoretical one neutron separation energies for
$^{11-18,20,22}$C are in agreement with the data. The calculated $S_{\rm n}$ is less
than 0 (-0.003 MeV) for the odd-$A$ nucleus $^{19}$C which is bound from experiment.
While for the experimentally unbound nucleus $^{21}$C, the predicted value of $S_{\rm
n}$ is positive. From the neutron-deficient side to the neutron drip line, excellent
agreement has been achieved for the nitrogen isotopes. Just as the other relativistic
mean field approaches, the RCHB calculations overestimate binding for $^{25,26}$O
which are unstable in experiment. For fluorine isotopes, the $S_{\rm n}$ in
$^{17,26-29}$F are overestimated in contrast with the underestimated one in $^{18}$F.
The neutron drip line nucleus is predicted as $^{30}$F. In general, the RCHB theory
reproduces the $S_{\rm n}$ data well considering that this is a microscopic and almost
parameter free model. There are some discrepancies between the calculations and the
empirical values for some of the studied isotopes. This may be due to  deformation
effects, which has been neglected here.

\begin{figure}
\centerline{\epsfig{figure=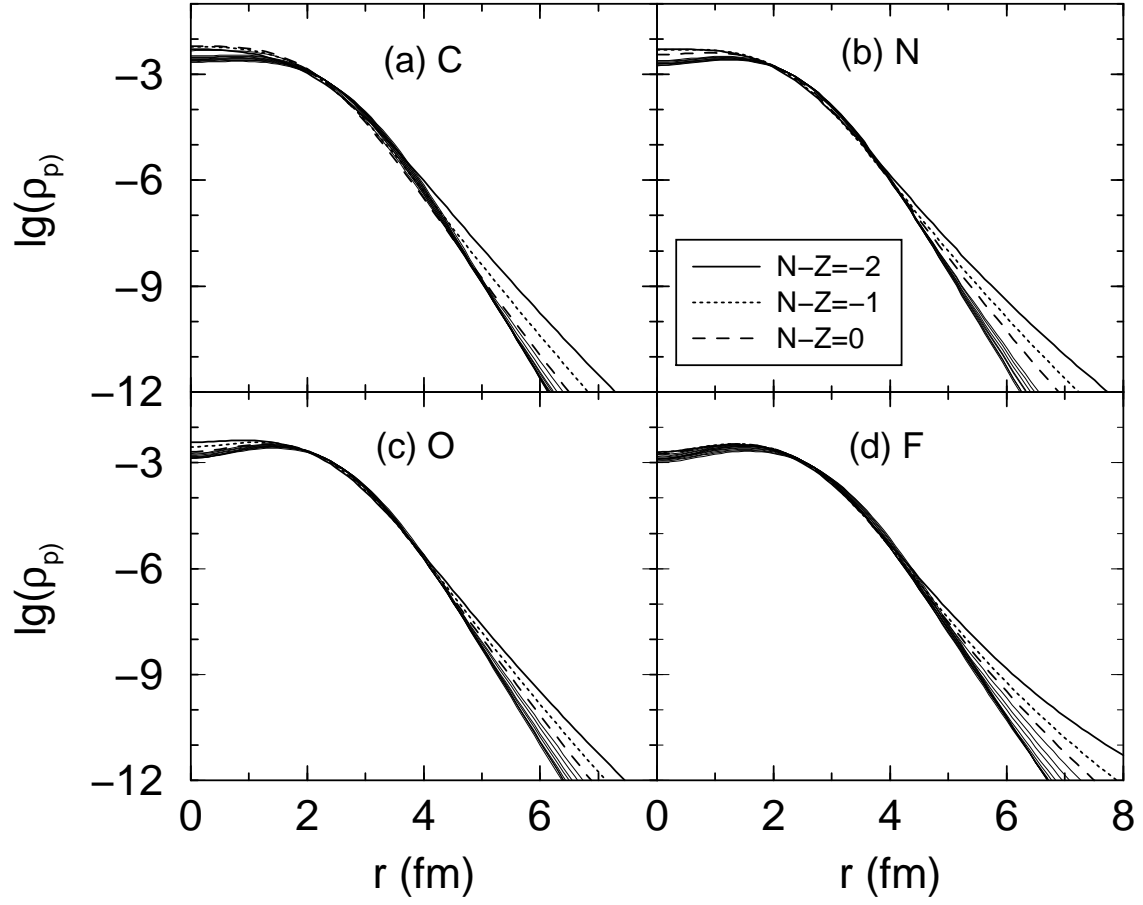,width=15.cm}}
 \vspace*{20pt}
 \caption{The proton density distributions
 predicted by RCHB for the nuclei
 $^{10-22}$C, $^{12-24}$N, $^{14-26}$O and $^{16-25}$F
 in logarithm scale.}
 \label{fig.d}
\end{figure}

The proton density distributions predicted by RCHB for the nuclei
$^{10-22}$C,$^{12-24}$N,$^{14-26}$O and $^{16-25}$F are given in
Fig.\ref{fig.d} in logarithm scale. The change in the density
distributions for each isotopes chain in Fig.\ref{fig.d} occurs
only at the tail or in the center part as the proton number is
constant. Because the density must be multiplied by a factor
$4\pi r^2$ before the integration in order to give proton number
or radii, the large change in the center does not matter very
much. What important is the density distribution in the tail
part. Compared with the neutron-rich isotopes, the proton
distribution with less $N$ has higher density at the center, lower
density in the middle part ( $2.5 < r < 4.5 $fm ), a larger tail
in the outer part ( $r > 4.5 $ fm ) which gives rise to the
increase of $r_{\rm p}$ and $\sigma_{\rm cc}$ for the proton rich
nuclei as will be seen in the following.

\begin{figure}
\centerline{\epsfig{figure=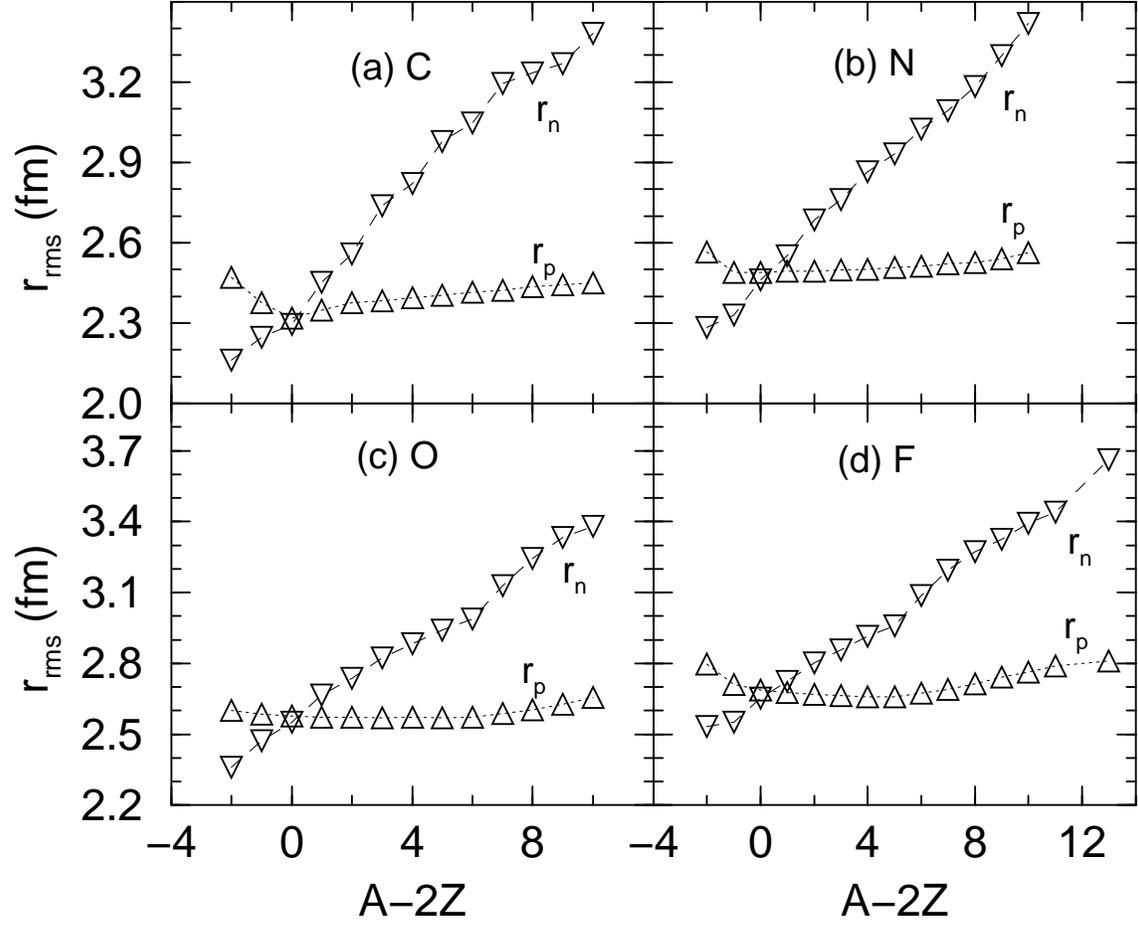,width=15.cm}}
\vspace*{23pt}
\caption{The neutron and proton rms
 radii predicted by RCHB for the nuclei $^{10-22}$C,
 $^{12-24}$N,$^{14-26}$O and $^{16-25}$F.}
 \label{fig.r}
\end{figure}

The neutron and proton rms radii predicted by RCHB for the nuclei
$^{10-22}$C, $^{12-24}$N,$^{14-26}$O and $^{16-25}$F are given in
Fig.\ref{fig.r}. The neutron radii for nuclei in each isotope
chain increase steadily. While the corresponding proton radii
remains almost constant with neutron number for nuclei in each
isotope chain except for the proton rich ones.

\begin{figure}
\centerline{\epsfig{figure=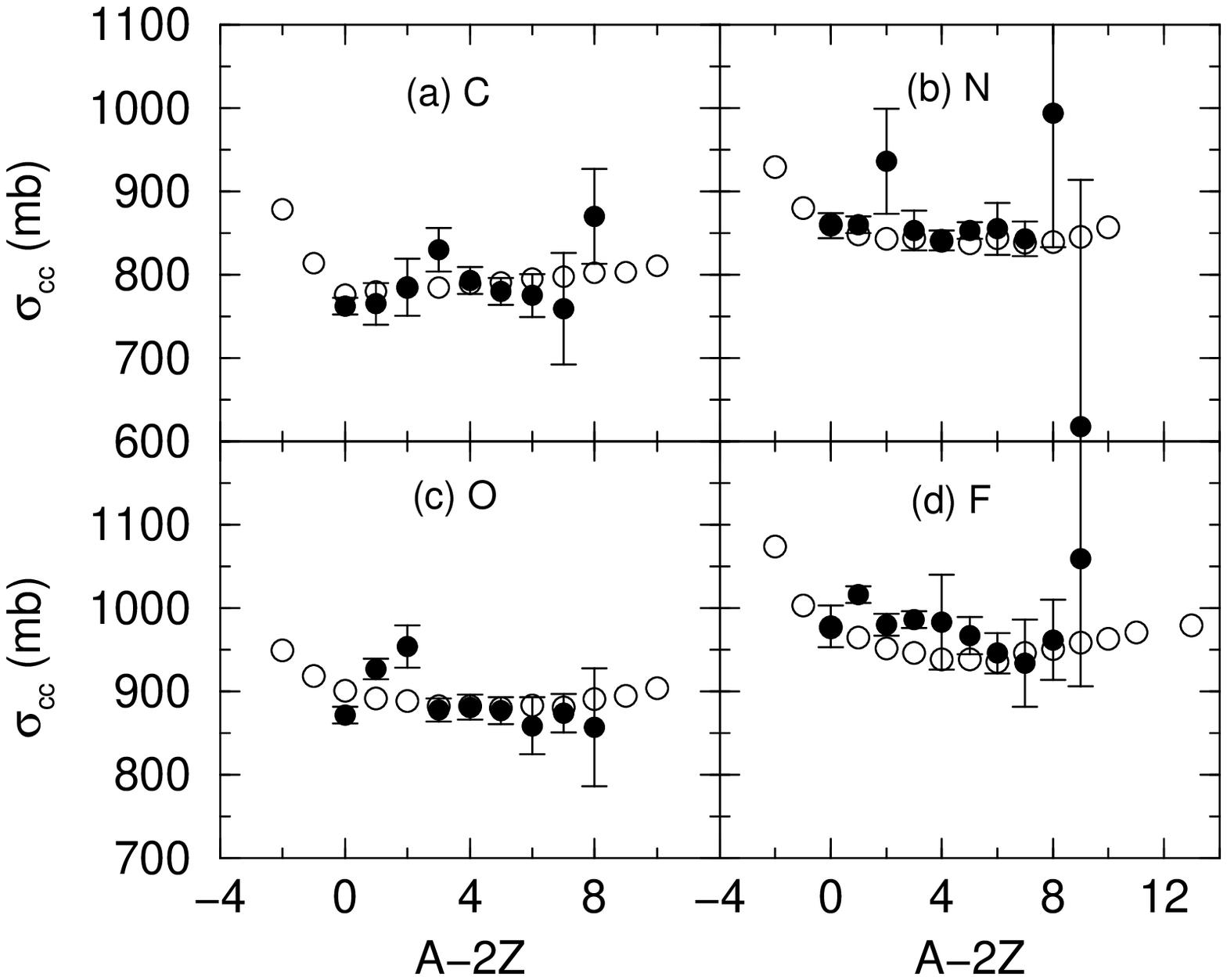,width=15.cm}} \vspace*{23pt}
\caption{The total charge-changing
 cross sections $\sigma_{\rm cc}$ of the nuclei
 $^{10-22}$C,$^{12-24}$N,$^{14-26}$O and $^{16-25}$F
 on a carbon target at relativistic energy.
 The open circles are the result of RCHB and the available
 experimental data are given by solid circles with their error-bars.}
 \label{fig.c}
\end{figure}

To compare the charge-changing cross sections $\sigma_{\rm cc}$
directly with experimental measured values, the densities
$\rho_{\rm p}(r)$ of the target $^{12}$C and the C, N, O and F
isotopes obtained from RCHB ( see Fig.\ref{fig.d} ) were used.
The cross sections were calculated in Glauber model by using the
free nucleon-nucleon cross section \cite{Ray.79} for the proton
and neutron respectively. The total charge-changing cross
sections $\sigma_{\rm cc}$ of the nuclei
$^{10-22}$C,$^{12-24}$N,$^{14-26}$O and $^{16-25}$F on a carbon
target at relativistic energy are given in Fig. \ref{fig.c}. The
open circles are the result of RCHB combined with the Glauber
Model and the available experimental data \cite{Chu.00} are given
by solid circles with their error-bars. The agreement between the
calculated results and measured ones are fine.

The charge changing cross sections change only slightly with the
neutron number except for proton-rich nuclei, which means the
proton density plays important role in determining the
charge-changing cross sections $\sigma_{\rm cc}$. A gradual
increase of the cross section has been observed towards the
neutron drip line. However, the big error bars of the data can not
help to conclude anything here yet. It is shown clearly that the
RCHB theory, when combined with the Glauber model, can provide
reliable description for not only interaction cross section but
also charge changing cross section. From comparison of this figure
and Fig.\ref{fig.r}, we can find similar trends of variations of
proton radii and of charge changing cross sections for each
isotope chain which implies again that the proton density plays
important role in determining the charge-changing cross sections.

Summarizing our investigations, the ground state properties for
C, N, O and F isotopes have been systematically studied with a
microscopic model --- the RCHB theory, where the pairing and
blocking effect have been treated self-consistently. The
calculated one neutron separation energies $S_{\rm n}$ are in
good agreement with the experimental values available with some
exceptions due to deformation effects which is not included in
the present study. A Glauber model calculation for the total
charge-changing cross section has been carried out with the
density obtained from the RCHB theory. A good agreement was
obtained with the measured cross sections for $^{12}$C as a
target. Another important conclusion here is that, contrary to
the usual impression, the proton density distribution is less
sensitive to the proton and neutron ratio. Instead it is almost
unchanged from stability to the neutron drip-line. The influence
of the deformation, which is neglected in the present
investigation, is also interesting to us, more extensive study by
extending the present study to deformed cases are in progress.

This work was partly supported by the Major State Basic Research
Development Program Under Contract Number G2000077407 and the
National Natural Science Foundation of China under Grant No.
10025522, 19847002 and 19935030.


\clearpage

\begin{references}
\bibitem{THH.85b} I. Tanihata,
 Prog. Part. Nucl. Phys.,{\bf 35}(1995) 505.
\bibitem{HJJ.95} P.G. Hansen, A.S. Jensen, and B. Jonson,
 Ann. Rev. Nucl. Part. Sci. {\bf 45} (1995) 591
\bibitem{Suz.95} T. Suzuki, et al.,
 Phys. Rev. Lett. {\bf 75} (1995) 3241
\bibitem{MTY.97} J. Meng, I. Tanihata and S. Yamaji,
 Phys. Lett. {\bf B419} (1998) 1.
\bibitem{Chu.00} L.V. Chulkov, et al.,
 Nucl. Phys. {\bf A674} (2000) 330
\bibitem{ME.98} J. Meng,
 Nucl. Phys. {\bf A635} (1998) 3.
\bibitem{MR.96} J. Meng, and P. Ring,
 Phys. Rev. Lett. {\bf 77} (1996) 3963.
\bibitem{MR.98} J. Meng, and P. Ring,
 Phys. Rev. Lett. {\bf 80} (1998) 460.
\bibitem{SW86} B.D. Serot and J.D. Walecka,
 Adv. Nucl. Phys. {\bf 16}, 1 (1986).
\bibitem{RE89} P-G. Reinhard,
 Rep. Prog. Phys. {\bf 52}, 439 (1989).
\bibitem{RI96} P. Ring,
 Prog. Part. Nucl. Phys. {\bf 37}, 193 (1996).
\bibitem{MSY.98} J. Meng, K. Sugawara-Tanabe, S. Yamaji, P. Ring and A. Arima,
 Phys. Rev. C58 (1998) R628.
\bibitem{Meng993} J. Meng, K. Sugawara-Tanabe, S. Yamaji and A. Arima,
 Phys. Rev. C 59 (1999) 154-163.
\bibitem{Meng992} J. Meng, and I. Tanihata,
 Nucl. Phys. A 650 (1999) 176-196.
\bibitem{Meng00} J. Meng, and N. Takigawa,
 Phys. Rev. C61 (2000) 064319.
\bibitem{RS.80} P. Ring and P. Schuck,
 {\it The Nuclear Many-body Problem\/}, Springer Verlag, Heidelberg (1980)
\bibitem{SNR.93} M. Sharma, M. Nagarajan and P. Ring,
 Phys. Lett. {\bf B312} (1993) 377
\bibitem{AUD.93}G. Audi, and A.H. Wapstra,
 Nucl. Phys. {\bf A565}  (1993) 1
\bibitem{Ray.79} L. Ray,
 Phys. Rev. {\bf C} 20 ( 1979 ) 1857
\end{references}
\end{document}